# Metal-insulator and insulator-insulator transitions in the quarter-filled band organic conductors

K.C. Ung, S. Mazumdar, and D.Toussaint
*Department of Physics, University of Arizona,*
*Tucson, AZ 85721, USA*

## Abstract

The theory of the $2k_F$ and $4k_F$ instabilities in quarter-filled band organic conductors is revisited. The phase angles of the $2k_F$ bond and charge density waves are shown to change as electron correlation is turned on, and this switching of the phase angle is critical for understanding the bond distortion patterns in the real materials. Intersite Coulomb interactions in the real materials must be nonzero but less than a critical value. Both intersite and intrasite charge density waves are destabilized in the quasi-two-dimensional regime for realistic parameters, thus explaining the weakening of these phases in the superconducting materials.

Conducting organic charge-transfer solids are of interest as highly doped Mott-Hubbard semiconductors. Nearly forty of these materials are superconducting, and evidence exists for strong Coulomb interactions among the fermions in these systems [1]. For a complete understanding of their normal state, it is essential that the spatial broken symmetries in the quasi-one-dimensional(quasi-one-d) nonsuperconducting conductors be precisely understood. In addition to the usual $2k_F$ Peierls instability, many of the organic conductors exhibit a $4k_F$ instability ($k_F$ is the one-electron Fermi wavevector) [2]. In spite of considerable theoretical work on the origin of the $4k_F$ instability, important issues involving distortions in the real materials and the parameter space in which the materials lie remain largely unresolved. This is primarily due to the limitations of calculations of susceptibilities [3–5] that do not measure actual distortion patterns or charge densities. We give here a considerably more detailed picture of the transitions in the actual experimental systems. The present picture is shown to have significant implications for the quasi-2-d superconductors, in which these transitions are absent or weakened.

We will limit ourselves to $\rho = 0.5$, where $\rho$ is the number of electrons per site. For this commensurate case, the bond order wave (BOW), with periodic modulation of the intersite distances, and the charge density wave (CDW), with modulation of the intrasite charge densities, are distinct. Each of these can have periodicities $2k_F$ (period 4) and $4k_F$ (period 2). Furthermore, each period 4 density wave can occur in two forms, corresponding to different phase angles. In Table 1 we show the various period 4 and period 2 patterns. For zero Coulomb interactions, the ground state consists of coexisting $2k_F$ BOW1 and CDW1 [6]. Table 1 also shows the experimentally observed bond distortion pattern [7] in $MEM(TCNQ)_2$ below the $2k_F$ transition. Our discussions are not limited to $MEM(TCNQ)_2$, rather, this is one system whose bond distortion pattern is precisely known.

Existing theories [3–5,8–10] do not explain several of the mysteries in the real materials. Observation of signatures of electron-molecular vibration (e-mv) coupling below the $4k_F$ transition [11] has in the past led to the interpretation of the $4k_F$ phase as a CDW [9,10]. This would imply (qualitatively speaking) alternate occupancies of sites by electrons (see Table 1, row 1, column 3). Previous work has shown that the $4k_F$ and $2k_F$ CDWs do not coexist [4]. Thus in such a case, for the $2k_F$ transition to occur on the same chain, the $2k_F$ phase can only be a spin density wave (SDW) or a spin-Peierls state. We discount the possibility of the SDW, as long range SDW in one dimension is not possible. The spin-Peierls state that can accompany the $4k_F$ CDW is the BOW1 state of Table 1, in which the distances between the "occupied" sites are alternating. The *experimental* bond tetramerized phase, shown in the last column of Table 1, is, however, different from BOW1, and therefore we conclude that the $4k_F$ cannot be a CDW to begin with. The $4k_F$ phase can also be a BOW (Table 1, column 3, second row), and indeed in this case peaks in the wavevector dependent charge-transfer susceptibility are found both at $2k_F$ and $4k_F$ within the extended Hubbard model [4]. What is still not clear, however, is how the experimental bond distortion pattern is obtained below the $2k_F$ transition, as no combination of the $4k_F$ BOW and the $2k_F$ BOW1 (which dominates over the BOW2 for noninteracting electrons) can give the observed distortion pattern. The strong role of the e-mv coupling [11] is also diffcult to understand if $4k_F$ is a BOW.

We show in the present work that these problems can be resolved if, (a) phonons are included explicitly in the calculations, and (b) actual bond distortion patterns and charge modulations are measured. The most significant findings of our calculations are: (i) the phase angles of the lowest energy period 4 structures change with nonzero Coulomb interaction. This switching of the phase angles is central to the detailed understanding of the $4k_F$ and the $2k_F$ transitions. (ii) Exactly as the $2k_F$ BOW1 and $2k_F$ CDW1 coexist for all e-mv and electron-acoustic phonon (e-ph) couplings [6], the $2k_F$ BOW2 and and the $2k_F$ CDW2 also coexist. (iii) The absolute ground state, for the intersite Coulomb interaction less than a critical value, has a $4k_F$ BOW component, in addition to the BOW2 and CDW2 components. The experimental lowest energy bond distortion pattern now emerges naturally as a combination of the period 2 BOW and the period 4 BOW2. (iv) Only for large intersite Coulomb interaction the ground state is a $4k_F$ CDW, which does not describe the experimental systems.

The Hamiltonian we consider is,

$$
\begin{aligned}
H= & \sum_{i\sigma}(t-\alpha(u_{i+1}-u_i))\left(c^\dagger_{i\sigma}c_{i+1,\sigma}+c^\dagger_{i+1,\sigma}c_{i\sigma}\right) \\
& +U\sum_i n_{i\uparrow}n_{i\downarrow}+V\sum_i n_i n_{i+1} \\
& +\frac{K_1^2}{2}\sum_i(u_{i+1}-u_i)^2+\beta\sum_i n_i v_i+\frac{K_2^2}{2}\sum_i v_i^2
\end{aligned}
\tag{1}
$$

Here $c^\dagger_{i\sigma}$ creates an electron of spin $\sigma$ at site $i$, $n_{i\sigma}=c^\dagger_{i\sigma}c_{i\sigma}$, $n_i=\sum_\sigma n_{i\sigma}$, $t$ is the one-electron hopping integral, and $U$ and $V$ the on-site and nearest neighbor Coulomb repulsions. The $u_i$ are the displacements of the ith molecular units from their equilibrium positions, $v_i$ corresponds to an internal molecular mode, $\alpha$ and $\beta$ the e-ph and e-mv coupling constants, and $K_1$ and $K_2$ are the corresponding elastic constants.

Exact numerical calculations were done for finite rings of 8, 12 and 16 sites for nonzero $U$ and $V$, both with and without the restriction of constant overall ring size. We first show the results for the case where the overall system size is constrained. We begin with $\beta=0$ and initially assume the functional form for $u_j$ that is valid in the $U=V=0$ limit, viz., $u_j=u_0 cos(2k_F j-\theta)$. The calculations are done for both $\theta=0$ and $\theta=\pi/4$, corresponding to the BOW1 and BOW2. The $2k_F$ BOW is weakened by $U$. However, the more significant result is the crossover from $\theta=0$ to $\theta=\pi/4$, corresponding to a transition from $2k_F$ BOW1 to $2k_F$ BOW2, for nonzero $U$. This crossover in the phase of the $2k_F$ BOW, which we show is critical for a complete understanding, is shown in Fig. 1(a).

In addition to causing a change in $\theta$, Coulomb interactions also add a $4k_F$-component to the ground state wavefunction. We repeat the calculation of the ground state energies with the Ansatz $u_j=u_0[r_2 cos(2k_F j-\theta_2)+r_4(cos 4k_F j-\theta_4)]$, where $r_2+r_4=1$. Numerical calculations indicate that $\theta_2=\pi/4$ and $\theta_4=0$. The absolute lowest energy within correlated models occurs for $r_4\neq 0$. Furthermore, $r_4$ increases monotonically with $U$, and for fixed $U$ increases rapidly with $V$, provided $V$ is less than a critical value $V_c(U)$. We discuss the issue of $V_c$ below. The effects of $U$ and $V$ on $r_4$ are summarized in Fig. 1(b).

We now emphasize the following point. The experimentally observed bond distortion pattern (last column of Table 1) is *exactly* what is obtained with our Ansatz for $r_4>0.41$. We conclude then that the crossover in $\theta$ and nonzero $r_4>0.41$ are both necessary to explain the experimental ground state bond distortion. Calculations of susceptibilities [3–5] do not distinguish between BOW1 and BOW2, and also cannot give the exact bond distortion pattern.

The issue of $V_c$ can be understood as follows. In the limit of $U \to \infty$ and $\beta = 0$, Eq. 1 reduces to the half-filled band spinless fermion Hamiltonian,

$$H = \sum_i (t - \alpha(u_{i+1} - u_i)) \left(a_i^\dagger a_{i+1} + a_{i+1}^\dagger a_i\right) + \frac{K_1^2}{2} \sum_i (u_{i+1} - u_i)^2 \quad (2)$$

Transition to a CDW with equal bond lengths occurs for $V > V_c = 2t$ and small $\alpha$. For $V < V_c$ a dimerized BOW appears for infinitesimal $\alpha$, with the bond dimerization increasing with $V$ [12]. Numerical simulations in the limit of $\alpha = 0$ indicate that even for finite U, a CDW appears only above a $V_c(U)$, which is now slightly larger than 2t [13]. The increase of $r_4$ with $V$ within our model is then expected for $V < V_c(U)$. Thus the observed periodicity in the last column of Table 1 indicates that experimental materials are in the $V < V_c$ regime.

The above results were confirmed by a second set of calculations. Instead of using the size constraint and the Ansatz for $u_i$, the ground state energy was minimized with respect to a position dependent $t_i$ using the Lanczos method in an iterative manner to find the final $t_i$. These calculations also find that for $U = V = 0$ the BOW1 is lower in energy, but for nonzero $U$ the BOW2 is favored.

As in the previous case, the absolute ground state is once again found to be a composite of the $4k_F$ BOW and the $2k_F$ BOW2. To display the distortion patterns we compute the Fourier transform $T(q)$ of the final hopping parameters $t_i$ which minimize the energy. Period four distortions appear as a peak in the Fourier transform at momentum $2k_F$, and the difference between the BOW1 and the BOW2 patterns is the phase of this Fourier component. Similarly the period 2 component of the distortion appears at momentum $4k_F$. Crossover occurs from a dominant $2k_F$ to a dominant $4k_F$ when $T(2k_F) = T(4k_F)$. Our results are summarized in Fig. 2(a) and (b), where the $2k_F$ curves correpond to the BOW2 phase, which is lower in energy. As in Fig. 1(b), here also nonzero $U$ enhances the $4k_F$ contribution to the wavefunction, but the $4k_F$ dominates over the $2k_F$ only for nonzero moderate $V$. Once the $4k_F$ contribution to the wavefunction dominates over the $2k_F$, the overall bond distortion pattern is exactly the same as the experimental pattern.

In the above we have discussed only the BOW. The experimental signature of e-mv coupling [11] requires also a CDW. Our numerical work finds that exactly as the BOW1 coexists with CDW1 [6], the BOW2, and the composite ground state of BOW2 and the $4k_F$ BOW, coexist with the CDW2. The CDW2 is confirmed by monitoring the actual site charge densities. As expected, the$2k_F$ CDW2 vanishes for $V > V_c(U)$, where the $4k_F$ CDW, not relevant for the experimental systems, appears.

The above results give a complete picture of the $2k_F$ and $4k_F$ instabilities in the real materials. The free energy of the experimental system is dominated by high spin states at high temperatures, whose electronic behavior is similar to that of the spinless fermion Hamiltonian of Eq. 2. For $V < V_c$ the lattice dimerizes to the $4k_F$ BOW below the $4k_F$ transition temperature. At still lower temperatures, the free energy is dominated by low spin states, and the behavior now would resemble that of the ground state of Eq.1. Dimerization of the dimerized lattice now takes place, and the overall bond distortion pattern resembles that in the last column of Table 1. Coexistence of these states with the $2k_F$ CDW2 indicates that the overall ground state is a composite of the $4k_F$ BOW, $2k_F$ BOW2, and $2k_F$ CDW2. Signatures of e-mv coupling [11] are then expected for any nonzero $\beta$.

Although the present work has focused on the (quasi-1-d) nonsuperconductors, our results have strong implications for the quasi-2-d superconducting TMTSF and the BEDT-TTF based materials. The tendency to have a phase transitions discussed here is considerably weakened in the superconductors [1]. In the quasi-2-d regime, the $4k_F$ CDW with alternate occupied sites is still possible. However, our demonstration that organic conductors have $V < V_c$ would explain the absence of this CDW transition. On the other hand, the absence of the BOW transition is a result of two-dimensionality. The BOW transition here is analogous to the spin-Peierls transition in $\rho = 1$, which is weakened in 2-d, and gives way to antiferromagnetism [14]. The CDW2, which we find here, is a result of the BOWs, and is thus also expected to vanish for sufficiently strong two-dimensional electron hopping. The only remaining possible transition is then to the $2k_F$ SDW, and it has been argued [15] that for $\rho$ away from 1 the SDW requires a minimal 2-d hopping, but then gradually weakens as the extent of two-dimensionality increases. This would be supported by the occurrence of a spin-Peierls transition in the quasi-1-d TMTT materials, the occurrence of SDW in the weakly two-dimensional TMTSF, its vanishing under pressure, and the absence of both spin-Peierle and SDW transitions in the less anisotropic BEDT-TTF. Thus all possible spatial broken symmetries are considerably weakened in $\rho = 0.5$ in the quasi-2-d regime for $V < V_c$. Whether or not superconductivity is related to the suppression of spatial broken symmetries is an intriguing question.

Some of these calculations were done on the iPSC/860 and Paragon at the San Diego Supercomputer Center. This work was supported by DOE grant DE-FG02–85ER–40213.

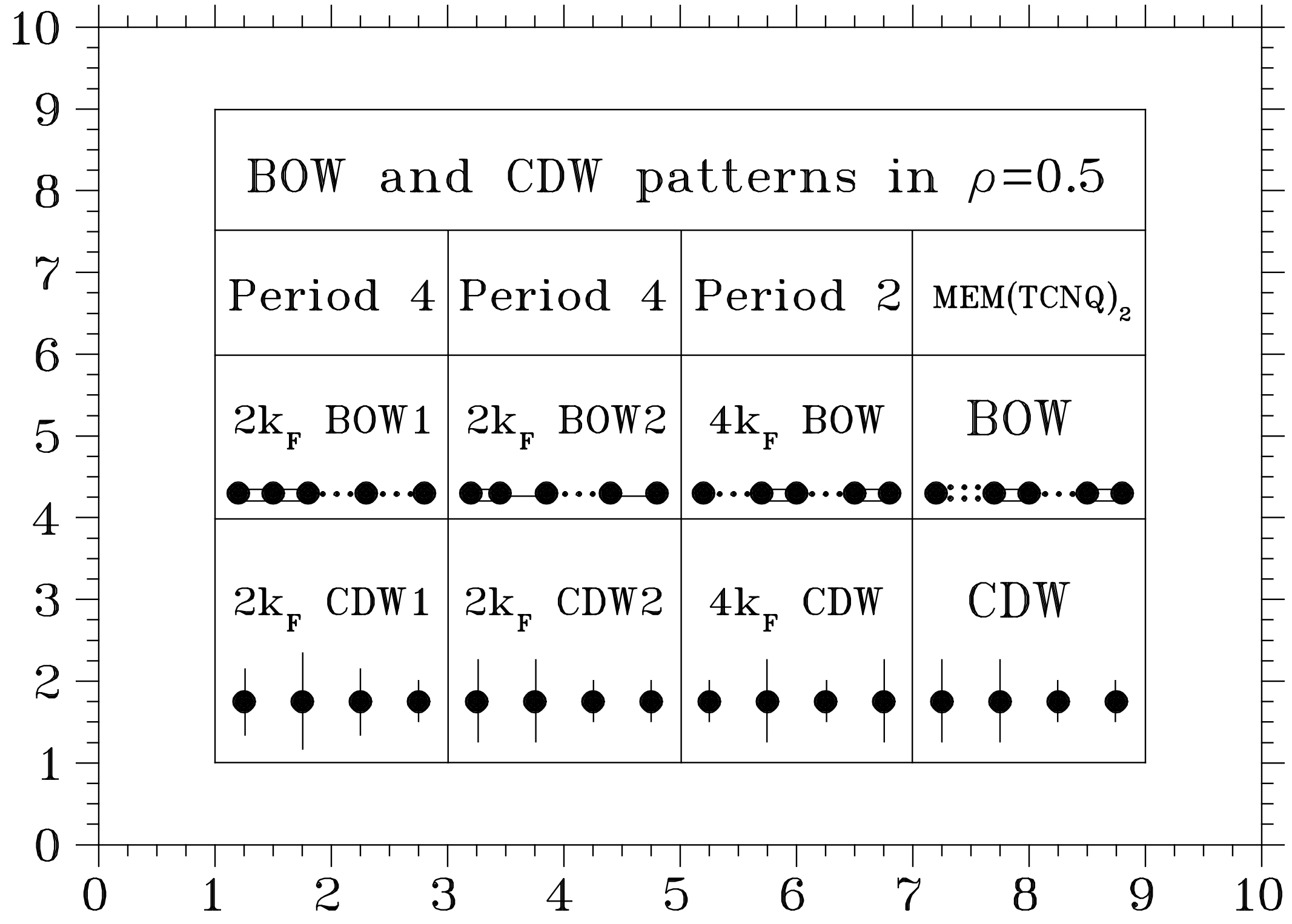

TABLE I. The possible bond distortions and charge modulations in a $\rho = 0.5$ chain. The double and dotted bonds correspond to bond lengths shorter and longer than the average (undistorted) bond length, while the single bonds correspond to undistorted bonds. In the last column we have given the bond distortion pattern in $MEM(TCNQ)_2$ below the $2k_F$ transition. Here the double dotted bond is not as long as the single dotted bond. In case of the CDWs, the lengths of the vertical bars on the sites correspond to the charge densities.

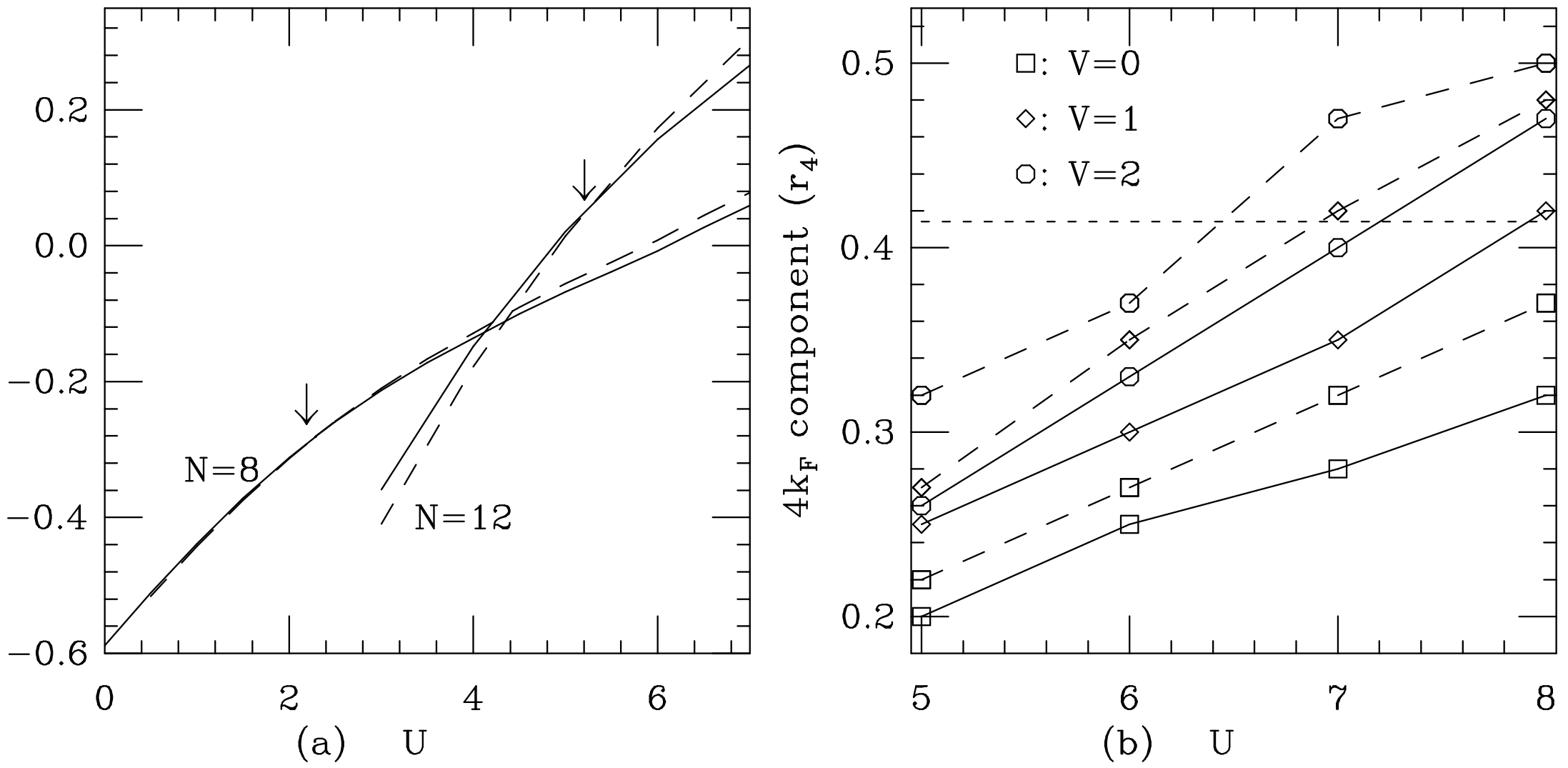

FIG. 1. The gain in total energy vs. $U$ for $2k_F$ bond distortions with two different phase angles for $N = 8$ and 12. (a) ($N$ is the length of the system.) The solid line in each case corresponds to the BOW1 and the dashed line to the BOW2. The arrows indicate the crossing of the BOW1 and BOW2 energies. (b) The relative weight $r_4$ of the $4k_F$ component of the ground state wavefunction as a function of $U$ and $V$. The dashed and solid curves correspond to $N = 8$ and 12 respectively. For $r_4 > 0.41$ the bond distortion pattern is that observed in MEM(TCNQ) below the $2k_F$ transition.

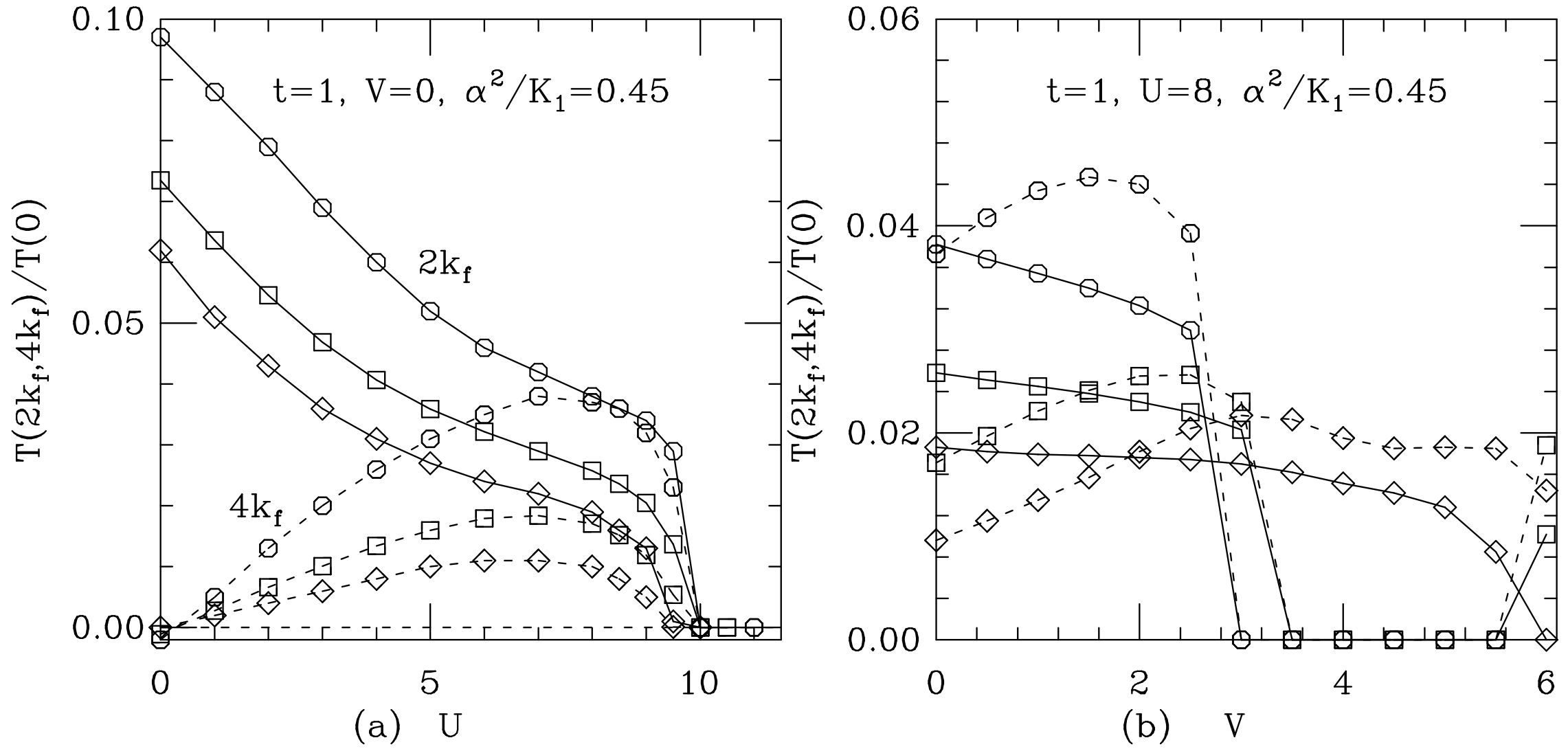

FIG. 2. The $2k_F$ (solid lines) and $4k_F$ (dashed lines) components of the hopping parameters $t_i$ at $V = 0$ and $\lambda = \alpha^2/K_1 = 0.45$. (a) Results are shown for $N = 8$ (circles) and $N = 16$ (diamonds) with periodic boundary conditions (circles), and for $N = 12$ with antiperiodic boundary conditions (squares). In all cases we see that the $4k_F$ component increases with $U$ until at very large $U$ the distortion vanishes completely. Note that, $U$ alone is insufficient to drive the $4k_F$ component larger than the $2k_F$ component, which is the condition for the pattern to change from "medium-short-medium-long" to "short-medium-short-long". (b): Same as (a) for fixed $U = 8$ and varying $V$. Note that $V$ enhances the $4k_F$ component, which can now be larger than the $2k_F$ component. The results for $L = 16$ at large $V$ may be an artifact of the small system size. At the end of this curve, at $V = 6$, the $L = 16$ distortion is purely $4k_F$, or period two. However, the wave function at this point has period of four, changing sign under translation by two lattice sites. For the undistorted lattice, this corresponds to a ground state with nonzero momentum.